\documentclass[aps,prl,noshowpacs,noshowkeys,floatfix,superscriptaddress,singlecolum,nofootinbib]{revtex4-2}

\def\apj{{ApJ}}                 
\def\aap{{A\&A}}                

\def\mnras{{MNRAS}}             
\def\prd{{Phys.~Rev.~D}}        
\def\prl{{Phys.~Rev.~Lett.}}    
\usepackage{amsmath,amstext}
\usepackage[T1]{fontenc}
\usepackage{amssymb}
\usepackage{graphicx}
\usepackage{ae,aecompl}

\usepackage{hyperref}
\usepackage{amsmath}
\usepackage{amssymb}
\usepackage{mathtools}
\usepackage{bm}
\usepackage{cleveref}
\usepackage{tensor}
\usepackage{braket}
\usepackage{enumitem}
\usepackage{mhchem}
\usepackage{cancel}
\usepackage{amsthm}
\usepackage{upgreek}
\usepackage{mathrsfs}
\usepackage{pgfplots}
\usepackage{tikz}
\usetikzlibrary{external}
\usepackage{color}

\newcommand{\spc}{\quad \quad \quad}

\def\be{\begin{equation}}
\def\ee{\end{equation}}
\def\beq{\begin{eqnarray}}
\def\eeq{\end{eqnarray}}

\theoremstyle{definition}

\theoremstyle{theorem}

\begin{document}
\title{Relativistic Bulk Rheology: From Neutron Star Mergers to Viscous Cosmology}
\author{L.~Gavassino}
\email{lorenzo.gavassino@vanderbilt.edu}
\affiliation{Department of Mathematics, Vanderbilt University, Nashville, TN 37211, USA}

\author{Jorge Noronha}
\email{jn0508@illinois.edu}
\affiliation{Illinois Center for Advanced Studies of the Universe \& Department of Physics, 
University of Illinois at Urbana-Champaign, Urbana, IL, 61801, USA}

\begin{abstract}
We develop the first causal and stable theory of a bulk-viscous relativistic pseudoplastic (or dilatant) fluid. This new formalism brings to light the rheological properties of several relativistic physical systems. Neutron star collisions can behave as a relativistic pseudoplastic material with viscous properties dictated by the non-conservation of lepton currents due to weak decay. Two-temperature relativistic plasmas, such as those surrounding supermassive galactic black holes, are predominantly pseudoplastic. Our framework can also be employed to construct novel viscous models for the evolution of the Universe with pseudoplastic or dilatant features.    
\end{abstract}

\maketitle

\noindent \textit{Introduction.} Rheology is the study of how matter responds to deformation \cite{malkin_book,Findley_book,Steffe_book}. Its main task is to determine relationships of the form $\Pi=\Pi[\theta]$, which express a certain stress component $\Pi$ as a functional of the deformation rate $\theta$ that caused such stress. Rheology emerged as a branch of fluid mechanics one century ago motivated by the observation that many real-world fluids (the so-called ``non-Newtonian fluids'' \cite{Tropea_book}) defy the canonical Navier-Stokes description \cite{LandauLifshitzFluids}. It was soon realized that all fluids are fundamentally non-Newtonian, and Navier-Stokes hydrodynamics is just the leading-order truncation of any (fluid-type) rheological relation $\Pi[\theta]$. Such truncation is allowed only at small spacetime gradients relative to some intrinsic scales of the fluid. Non-Newtonian corrections appear if such intrinsic scales are comparable to those associated with the variation of the hydrodynamic variables \cite{Frenkel_book,BAGGIOLI20201,BaggioliHolography}.

Since most engineering-related flows (e.g., the Couette flow \cite{landau6}) are incompressible, non-relativistic rheology mainly focuses on shear phenomena. However, in relativity, the situation is profoundly different. Many relativistic flows (e.g., Bjorken flow \cite{Bjorken1983} in the context of heavy-ion collisions) experience large expansion rates. More importantly, systems whose description relies on relativistic fluid dynamics, namely the quark-gluon plasma formed in heavy-ion collisions \cite{Romatschke:2017ejr}, neutron star merger simulations \cite{Rezzolla_Zanotti_book}, relativistic plasmas surrounding black holes \cite{EventHorizonTelescope:2022urf}, and viscous cosmology \cite{Brevik:2017msy}, require careful treatment of bulk (i.e., expansion-induced) viscosity, and are expected to explore dynamical regimes where the simple Navier-Stokes truncation \cite{LandauLifshitzFluids,EckartViscous} cannot be applied without displaying causality violation and unphysical instabilities \cite{Hiscock_Lindblom_instability_1985}. Therefore, one needs a causal and stable \emph{relativistic theory of bulk-viscous rheology} that allows us to express the bulk stress $\Pi$ in terms of the relativistic expansion rate $\theta=\nabla_\mu u^\mu$, where $u^\mu$ is the fluid's 4-velocity. Here, we set the foundations of such a theory and discuss a few relevant applications. This work sheds new light on open questions concerning viscous hydrodynamics in neutron star mergers \cite{AlfordBulk,BulkGavassino,Celora:2022nbp,CamelioSimulations2022,Most:2021zvc,Most:2022yhe}, heavy-ion collisions (e.g. concerning nonlinear causality \cite{Bemfica:2020xym,Plumberg:2021bme,Chiu:2021muk}, cavitation \cite{Torrieri:2008ip,Rajagopal:2009yw,Denicol:2015bpa,Byres:2019xld}, attractors \cite{Heller:2015dha,Romatschke:2017vte,Strickland:2017kux,Jankowski:2023fdz}), and cosmology \cite{Padmanabhan:1987dg,Hiscock:1991sp,Pavon:1990qf,Zakari:1993yhk,Maartens:1995wt,Zimdahl:1996ka,Fabris:2005ts,Colistete:2007xi,LiBarrow,Avelino,Hipolito1,Hipolito2,Gagnon,Piattella_et_al,Velten:2012uv,VeltenSchwarz,BG,Disconzi_Kephart_Scherrer_2015,DisconziKephartScherrerNew,
Cruz:2018psw,Cruz:2019uya,Brevik:2017msy}.   
\noindent \emph{Notation}: Our metric has signature $(-+++)$ and $c=k_B=1$. When convenient, the notation $\dot{X}=u^\mu \nabla_\mu X$ is also adopted.\\

\noindent \textit{Statement of the problem.} The simplest rheological model for fluids is the Navier-Stokes constitutive relation, $\Pi=-\zeta \theta$, where $\zeta>0$ is a linear susceptibility (independent from $\theta$), known as the bulk viscosity coefficient \cite{LandauLifshitzFluids}. This model follows\footnote{A choice of hydrodynamic frame is also made, see \cite{Kovtun:2012rj}.} from assuming a quasi-stationary process and a small $\theta$ (compared to some intrinsic scale of the system) expansion. Real-world flows may break both these assumptions. If the first assumption is violated, the fluid is called \emph{viscoelastic} \cite{Findley_book}. In a  \emph{pseudoplastic} fluid, the second assumption is violated \cite{Steffe_book}. The associated constitutive relations are respectively\footnote{Equation \eqref{two} is just the simplest model for viscoelasticity, known as the ``Maxwell model'' \cite{landau7,BaggioliHolography,GavassinoGENERIC2022}. Other more complicated expressions are known (e.g., the Burgers model \cite{Malek2018,GavassinoBurgers2023}). The most general time-dependent linear rheological relation takes the form $\Pi(t)=\int G(t{-}t')\theta(t')dt'$, where $G$ is some linear-response Green's function \cite{Denicol_Relaxation_2011}.}
\begin{flalign}
\noindent
\text{Viscoelastic:} && \label{two} \tau \dot{\Pi}+\Pi &= -\zeta \theta \;,&\\
\noindent
\text{Pseudoplastic:} && \label{three} \Pi &= -f(\theta) \;,
\end{flalign}
where above $\tau>0$ is a transport coefficient (the intrinsic scale), independent from $\theta$, while $f$ is an arbitrary function of $\theta$. Physically, a viscoelastic fluid is a material whose stress exhibits a delay in the response to time-dependent deformation rates. The term reflects the fact that, in the high-frequency limit, \eqref{two} reduces to Hooke's law of elasticity, $\dot{\Pi}\propto \theta$ \cite{landau7}. A pseudoplastic fluid is a material whose bulk viscosity coefficient, now defined as $\zeta=-\Pi/\theta$, changes as a function of the deformation rate. The term comes from the fact that, usually, such a $\zeta(\theta)$ decreases with $\theta$ so that the induced stress is relatively small at large deformation rates. As discussed below, both effects are present in neutron-star matter, QCD critical dynamics, relativistic plasmas surrounding black holes, and viscous cosmology.

The relativistic theory for viscoelasticity goes under the name of Israel-Stewart theory \cite{MIS-6,Hiscock_Lindblom_stability_1983}. Born as an approximation of kinetic theory \cite{Denicol2012Boltzmann}, its accuracy as an (almost \cite{Heller2014}) universal rheological model has been recently established systematically, both within a thermodynamic \cite{GavassinoUniversalityI,GavassinoUniversalityII2023} and a linear-response framework \cite{WagnerGavassino2023}. On the other hand, a relativistic theory for pseudoplasticity is still missing. A straightforward implementation of \eqref{three} in a relativistic hydrodynamic model would result in causality violation and ultraviolet instabilities. Even combining \eqref{two} and \eqref{three} into $\tau \dot{\Pi}+\Pi =-f(\theta)$, the resulting system of equations would not be quasilinear \cite{ChoquetBruhatGRBook}, leading to insurmountable difficulties in establishing causality and well-posedness of the initial value problem \cite{Disconzi:2023rtt}. Moreover, without a systematic procedure for computing $f(\theta)$ from microphysics, the applicability of rheological concepts in concrete relativistic systems remains hypothetical. Below, we present a simple mathematical procedure that allows one to rewrite previously existing frameworks in rheological form, automatically including viscoelasticity and pseudoelasticity. The resulting hydrodynamic theory can be easily proven causal, stable, strongly hyperbolic, and thermodynamically consistent. Furthermore, general formulas are given to compute the rheological transport coefficients directly from microphysics.\\

\noindent \textit{General modeling.} Our starting point is a simple observation: most of the relativistic systems where bulk viscosity is important, including our main systems of interest (neutron-star matter, QCD near the critical point, ionized plasmas, and cosmological fluids), can be modeled within the same framework, which we summarize below. 

We consider the case where four effective fields parameterize the macroscopic state of the fluid: $\{u^\mu,\rho,n,\phi \}$, representing the 4-velocity, the rest-frame energy density, the rest-frame baryon density, and a non-equilibrium excursion parameter, respectively. The first three fields are the usual fluid dynamical variables whose existence reflects a corresponding conservation law (momentum, energy, and baryon number). The scalar field $\phi$ is an observable that, due to the expansion of the fluid, is driven out of local equilibrium.  Usually, $\phi$ reflects the existence of a weakly broken conservation law\footnote{In response theory, a weakly broken conservation law is a gapped mode $\phi$ whose relaxation time $\tau_\phi$ is parametrically slow compared to the relaxation time $\tau_{\text{micro}}$ of all other gapped modes \cite{Grozdanov2019,GavassinoNonHydro2022}. Under this assumption, one can define a quasi-hydrodynamic regime, valid over the timescale $\tau_{\text{micro}} \ll t \lesssim \tau_\phi$, where the fluid exists in a quasi-equilibrium state, and $\phi$ plays the role of a generalized thermodynamic variable (see \cite{landau5} \S 118).}, e.g., a chemical reaction \cite{GavassinoFronntiers2021}. Since here we only focus on bulk viscosity, we assume that the stress-energy tensor $T^{\mu \nu}$ and the baryon current $n^\mu$ are isotropic in the local rest frame so that \cite{MTW_book}
\begin{equation}
    T^{\mu \nu} = (\rho+P)u^\mu u^\nu + Pg^{\mu \nu} \, , \spc n^\mu = nu^\mu \, ,
\end{equation}
where $P(\rho,n,\phi)$ is the \emph{non-equilibrium rheological pressure} and $g^{\mu\nu}$ is the (arbitrary) spacetime metric. The equations of motion of the fluid are, therefore,
\begin{equation}\label{sbam}
\begin{split}
u^\mu \nabla_\mu u_\nu={}& - (g\indices{^\mu _\nu}{+}u^\mu u_\nu) \dfrac{\nabla_\mu P}{\rho+P} \, , \\
   u^\mu \nabla_\mu\rho={}& -(\rho+P)\nabla_\mu u^\mu \, , \\
   u^\mu \nabla_\mu n={}& -n\nabla_\mu u^\mu \, , \\
   u^\mu \nabla_\mu \phi ={}& -K \nabla_\mu u^\mu -F \, . \\
\end{split}
\end{equation}
The first three equations are the conservation laws, $\nabla_\mu T^{\mu \nu}=0$ and $\nabla_\mu n^\mu=0$. The last equation models dissipation. Its structure directly follows from isotropy and the assumption that $\phi$ is the only non-equilibrium degree of freedom. The coefficient $K(\rho,n,\phi)$ is called the \emph{compressibility} and $F(\rho,n,\phi)$ the \emph{returning force}. Therefore, to completely specify the systems we are interested in, we only need to know how to express $\{P,K,F\}$ in terms of $\{\rho,n,\phi\}$.

Finally, the reader may note a similarity between \eqref{sbam} and the Hydro+ effective theory framework, which was introduced in Ref.\ \cite{Stephanov:2017ghc} to describe
a near-hydrodynamic system with an additional mode that is
parametrically slower than the other modes. This is not a coincidence, given the broad regime of applicability of Hydro+ \cite{Stephanov:2017ghc}. \\ 

\noindent \textit{Bulk rheology.} Let us recast \eqref{sbam} into a rheological model for bulk viscosity, which also accounts for pseudoplasticity. First, we note that the equilibrium value $\phi_{\text{eq}}(\rho,n)$ of the non-conserved mode $\phi$ (at given $\rho$ and $n$) can be computed from the relation $F\big(\rho,n,\phi_{\text{eq}}(\rho,n)\big)=0$ \cite{Stephanov:2017ghc}. This follows from the fact that the equilibrium state is stationary and non-deforming so that $0=u^\mu \nabla_\mu \phi=-F$ at equilibrium. Then, given the expression for the non-equilibrium pressure, $P(\rho,n,\phi)$, we can define the equilibrium pressure and the bulk scalar as
\begin{equation}\label{pI}
P_{\text{eq}}(\rho,n)= P\big(\rho,n,\phi_{\text{eq}}(\rho,n)\big),\qquad
\Pi(\rho,n,\phi)= P\big(\rho,n,\phi)-P\big(\rho,n,\phi_{\text{eq}}(\rho,n)\big),
\end{equation}
so that the energy-momentum tensor takes the usual bulk-viscous form
\begin{equation}
T^{\mu \nu}=(\rho+P_{\text{eq}}+\Pi)u^\mu u^\nu+(P_{\text{eq}}+\Pi)g^{\mu \nu}.
\end{equation}
Inverting the second equation of \eqref{pI}, we obtain a relation of the form $\phi=\phi(\rho,n,\Pi)$. Clearly, $\phi_\text{eq}(\rho,n)=\phi(\rho,n,0)$ and, since $F$ vanishes at equilibrium,  $F_\Pi=F/\Pi$ is non-singular at $\Pi=0$ as it converges to $\partial_\Pi F(\Pi=0)$. Expressing the last equation of \eqref{sbam} in terms of $\Pi$, we obtain a relation that resembles Israel-Stewart theory (though here the equation is valid also for large $\Pi/P_{\textrm{eq}}$): 
\begin{equation}
\tau_\Pi u^\mu \nabla_\mu \Pi +\Pi = -\zeta \nabla_\mu u^\mu ,
\label{seven}
\end{equation}
where $\tau_\Pi(\rho,n,\Pi)$ and $\zeta(\rho,n,\Pi)$ are now also rheological functions of $\Pi$, given by
\begin{equation}\label{dictionary}
\tau_\Pi = \dfrac{1}{F_\Pi} \dfrac{\partial \phi}{\partial \Pi}\bigg|_{\rho,n} , \qquad
\zeta = -\dfrac{1}{F_\Pi} \bigg[ (\rho{+}P_{\text{eq}}{+}\Pi)\dfrac{\partial \phi}{\partial \rho}\bigg|_{n,\Pi} \! \! \! +n \dfrac{\partial \phi}{\partial n}\bigg|_{\rho,\Pi} - K  \bigg]. 
\end{equation}
No approximation has been made above, meaning that  \eqref{seven} is mathematically equivalent to \eqref{sbam} (the equivalence also holds in curved spacetime).

Equation \eqref{seven} is the relativistic theory for bulk rheology we were looking for. Indeed, viscoelasticity is automatically accounted for by the relaxation time, and all the available formulations of Israel-Stewart bulk viscosity (e.g. DNMR \cite{Denicol2012Boltzmann}, or Hiscock-Linblom \cite{Hiscock_Lindblom_stability_1983}) correspond to particular choices of $\tau_\Pi(\Pi)$ and $\zeta(\Pi)$ (see Supplemental Material, Sec. III). However, the theory can also describe pseudoplasticity. In fact, suppose that the system has reached an attractor state \cite{Heller:2015dha}, where $\tau_\Pi u^\mu \nabla_\mu \Pi$ can be replaced with some function $h(\rho,n,\Pi,\nabla_\mu u^\mu)$. Then, we have an equation of the form $h(\Pi,\nabla_\mu u^\mu)+\Pi=-\zeta(\Pi)\nabla_\mu u^\mu$. If we isolate $\Pi$, we obtain a (fully nonlinear) expression $\Pi=-f(\nabla_\mu u^\mu)$, which can be interpreted as a late-time pseudoplastic constitutive relation. This further develops the idea of effective transport coefficients that encode the contribution from an infinite number of gradients, previously investigated in relativistic systems undergoing highly symmetric flows  \cite{Romatschke:2017vte,Denicol:2017lxn,Blaizot:2017ucy,Behtash:2018moe,Denicol:2020eij}. In this context, the results of \cite{Romatschke:2017vte} show that the hydrodynamic attractor found in kinetic and holographic systems undergoing Bjorken flow is in the pseudoplastic regime. We note, however, that the dependence of $\zeta$ with gradients discussed above holds, in different forms, for arbitrary flows (also in curved spacetime). Furthermore, causality and stability issues are automatically solved since \eqref{seven} gives rise to a fluid model that is thermodynamically consistent, symmetric hyperbolic, and causal in the fully nonlinear regime (see Supplemental Material, Sec. I).\\


\noindent \textit{Bulk-viscous rheology of neutron star mergers.} We now argue that the ultradense matter formed in neutron star mergers must have bulk-viscous rheological properties. Current state-of-the-art simulations of neutron star mergers (see, e.g., \cite{Most:2022yhe}) solve \eqref{sbam} coupled to Einstein's equations. Assuming $npe$ matter in the neutrino transparent regime, $\phi$ corresponds to the charge fraction $Y=n_e/n$ ($n_e$ is the electron density), which only changes by weak-interaction decays of neutrons and protons. This gives $K=0$ \cite{noto_rel,comer_langlois1994} and $F=\Gamma_\nu/n$, where $\Gamma_\nu$ are the weak-interaction rates, which include standard leakage schemes \cite{1996A&A...311..532R,2003MNRAS.342..673R} as well as direct and modified Urca net rates \cite{Yakovlev:2000jp,Alford:2021ogv}. In this case, $P = P(\rho,n,Y)$ can be determined directly from the underlying model for the equation of state away from beta equilibrium. Then, our ``dictionary relations" \eqref{dictionary} can be used to directly determine the transport coefficients for neutron-star matter far from beta equilibrium and define its rheological properties. Indeed, the first study of this kind has been carried out in \cite{Yang:2023ogo}, where our transport coefficients in \eqref{dictionary} were evaluated numerically using a realistic equation of state compatible with astrophysical constraints. Therefore, the exact mathematical equivalence between \eqref{sbam} and \eqref{seven}, established here, conclusively shows for the first time that the matter formed in neutron star collisions is intrinsically viscous, going beyond the linear response analyses of \cite{Gavassino:2020kwo,Celora:2022nbp,Camelio2022}. This general result is valid for arbitrary equations of state, rates, and
dynamical spacetimes. Whether or not such viscous effects can be measured using gravitational waves is still under debate \cite{Alford:2017rxf,Most:2021zvc, Hammond:2022uua,Most:2022yhe,Chabanov:2023blf,Espino:2023dei,Ripley:2023lsq}.\\

\noindent \textit{An analytical model.}
To further discuss the physics behind the mathematical mapping leading to \eqref{dictionary} and how pseudoplasticity affects the dynamics of the bulk stress, we consider below an oversimplified toy model for neutron-star matter, where the mapping can be carried out analytically. 
Inspired by the neutron-star models adopted in \cite{Camelio2022,CamelioSimulations2022}, we consider the following relations\footnote{We have set all dimensional constants to one for convenience. Furthermore, we have decided to neglect the temperature dependence of the pressure and the reaction rate to simplify the analysis.}:
\begin{equation}\label{quaccheri}
P=n^2 e^{-Y},\qquad 
F= (n^{-1/2}e^Y-1)C^{-1}, 
\end{equation}
where $C {>} 0$ is a constant. This choice of $P$ and $F$ reproduces some qualitative features that neutron-star matter is expected to have. For example, $P$ decreases with $Y$ at constant $n$ \cite{Yang:2023ogo}. 
Additionally, the equilibrium fraction $Y_{\text{eq}}{=}\ln\sqrt{n}$ (computed from the requirement that $F_{\text{eq}}=0$) increases with $n$, as predicted by nuclear models \cite{CamelioSimulations2022}. In the Supplementary Material (Sec. II), we verify that the resulting equations of motion are indeed causal, strongly hyperbolic \cite{rezzolla_book}, and thermodynamically consistent arbitrarily far from beta equilibrium, besides being covariantly stable \cite{GavassinoSuperluminal2021,GavassinoBounds2023}, both thermodynamically and hydrodynamically \cite{GavassinoGibbs2021,
GavassinoCausality2021,GavassinoStabilityCarter2022}. Using \eqref{pI}, we obtain the equilibrium pressure and the bulk stress: $P_{\text{eq}}= n^{3/2}$ and $\Pi = n^2 e^{-Y}-n^{3/2}$. From this, the expressions for the transport coefficients \eqref{dictionary} follow immediately: 
\begin{equation}\label{exact}
\tau_\Pi = C,\quad \quad
\zeta =\bigg[\dfrac{n^{3/2}}{2} +2\Pi \bigg]C .
\end{equation}
One can see that $\zeta$ exhibits a non-trivial dependence on $\Pi$. Note that \eqref{exact} holds for arbitrarily large values of $\Pi$.\\

\noindent\textit{Pseudoplasticity as an attractor.} We can show that neutron-star matter modeled by \eqref{exact} is pseudoplastic. To this end, let us solve \eqref{seven} along the worldline of a fluid element (i.e., along an integral curve of $u^\mu$), parameterized with the proper time $t$. Assuming a constant expansion rate $\nabla_\mu u^\mu$, and using the transport coefficients \eqref{exact}, we find
\begin{equation}\label{gruenz}
  \dfrac{\Pi(t)}{P_{\text{eq}}(t)} = A \, e^{-\big(1{+}\frac{1}{2}\tau_\Pi\nabla_\mu u^\mu\big)t/\tau_\Pi} - \dfrac{\frac{1}{2}\tau_\Pi\nabla_\mu u^\mu}{1+\frac{1}{2}\tau_\Pi\nabla_\mu u^\mu}  \, ,
\end{equation}
where $A$ is an integration constant. We see that if $\tau_\Pi\nabla_\mu u^\mu >-2$, then the system admits a late-time attractor (the solution with $A=0$), where $\Pi$ is a nonlinear function of $\nabla_\mu u^\mu$. We can express this  \emph{late-time constitutive relation} in Navier-Stokes form, $\Pi =-\zeta_{\text{res}} \nabla_\mu u^\mu$, by defining a resummed bulk viscosity coefficient (see Fig.\ \ref{figunzzuzzuz}):
\begin{figure}
\centering
\includegraphics[width=0.5\textwidth]{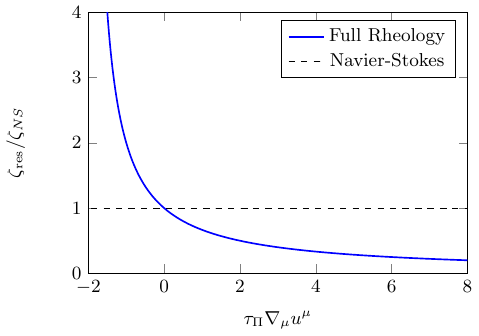}
\caption{Pseudoplastic features of our neutron-star matter toy model \eqref{quaccheri} (blue) compared with the Navier-Stokes prediction (dashed). The resummed bulk viscosity coefficient $\zeta_{\text{res}}$ (rescaled by $\zeta_{NS}=P_{\text{eq}}\tau_\Pi/2$) is plotted as a function of the expansion rate $\nabla_\mu u^\mu$ (in units of $\tau_\Pi^{-1}$).}\label{figunzzuzzuz}
\end{figure}
\begin{equation}
    \zeta_{\text{res}} = \dfrac{ \zeta_{NS} }{1{+}\frac{1}{2}\tau_\Pi \nabla_\mu u^\mu}  \, ,
\end{equation}
where $\zeta_{NS}=\zeta(\Pi{=}0)$ is the transport coefficient that would enter the Navier-Stokes model. As can be seen, if the fluid expands (i.e., $\nabla_\mu u^\mu {>}0$), the effective viscosity is reduced, and it tends to zero when $\nabla_\mu u^\mu \rightarrow +\infty$. This is the standard signature of pseudoplasticity. If, however, the fluid is compressed (i.e., $\nabla_\mu u^\mu {<}0$), then the effective viscosity becomes much larger than the Navier-Stokes viscosity. In rheology, this is the hallmark of dilatant behavior \cite{Steffe_book}. For $\tau_\Pi \nabla_\mu u^\mu <-2$, the general solution \eqref{gruenz} does not have a late-time attractor. \\ 

\noindent\textit{Impact of pseudoplasticity.} If a bulk-viscous fluid undergoes small quasi-periodic oscillations of frequency $\omega$ around thermodynamic equilibrium, the (approximate) damping time of the oscillation due to bulk dissipation is \cite{CamelioSimulations2022}
    $t_{\text{damp}} = 2c_s^2 (\rho+P_{\text{eq}})(1+\omega^2 \tau_\Pi^2)/\zeta \omega^2$,
where $c_s$ is the speed of sound. Keeping $\omega$ fixed, and treating $t_{\text{damp}}$ as a function of the reaction rate intensity $C^{-1}$ [see \eqref{quaccheri} and \eqref{exact}],  $t_{\text{damp}}$ has an absolute minimum when $\tau_\Pi=\omega^{-1}$. For this reason, it has been argued that in neutron star mergers, bulk viscosity should have the strongest impact in resonant regions \cite{AlfordRezzolla}, where the relaxation time scale of $\beta-$reactions equals the timescale of the hydrodynamic evolution. However, since such an estimate is carried out in the linear regime, it can only account for viscoelastic effects, and it completely neglects pseudoplasticity. Interestingly, we see from Fig.\  \ref{figunzzuzzuz} that pseudoplastic corrections are largest not at $\tau_\Pi \nabla_\mu u^\mu=1$ (as the resonance argument might suggest) but in the limit as $\tau_\Pi \rightarrow +\infty$, i.e. when dissipation is negligible. This explains why, in recent simulations of neutron-star migration \cite{CamelioSimulations2022} (which is a highly nonlinear process), rheological effects were shown to be the largest when beta reactions are suppressed, i.e., for $u^\mu \nabla_\mu Y \approx 0$.\\

\noindent\textit{Two-temperature plasmas.} Due to the high temperatures (around $10^{7}-10^{13}$ K) achieved in black hole accretion, hydrogen becomes fully ionized \cite{Shapiro_book}. Since the $ee$ and $pp$ inelastic collisions are much faster than $ep$ inelastic collisions \cite{bellan_2006}, the electron and proton gases achieve kinetic equilibrium at two different temperatures \cite{EventHorizonTelescope:2019pgp}. This allows us to rigorously model the plasma using the formalism presented in this work, where the additional non-equilibrium variable $\phi$ can be identified with the electron pressure. Considering  ultrarelativistic electrons and non-relativistic protons (with mass set to unity, for simplicity), one can straightforwardly derive the constitutive relations from thermodynamics (see Supplementary Material, Sec. IV):
\begin{equation}
    P=\dfrac{2}{3}(\rho-n)-\phi \, , \spc K=\dfrac{4}{3}\phi \, , \spc F=\dfrac{1}{C}\bigg[\phi -\dfrac{2}{9}(\rho-n) \bigg] \, ,
\end{equation}
where $C>0$ is a constant for simplicity. The exact bulk transport coefficients in the rheological representation are
\begin{equation}
  P_{\text{eq}}=\dfrac{4}{9}(\rho-n)\, , \spc \tau_\Pi = C \, , \spc \zeta = \dfrac{2C}{81}(\rho-n+63 \, \Pi) \, .    
\end{equation}
The equilibrium equation of state for the pressure has an adiabatic index $13/9$ \cite{Shapiro_book}, in agreement with simulations of M 87 jets \cite{Moscibroska2016}. Working out the late-time attractor of a plasma undergoing uniform expansion leads to the resummed bulk viscosity coefficient (see Fig.\ \ref{figunzzuzzuz2}):
\begin{figure}
\begin{center}
    \centering
    \includegraphics[width=0.5\textwidth]{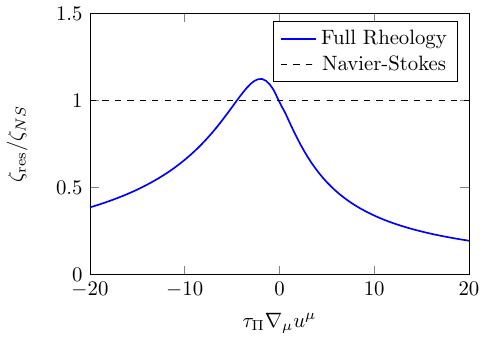}
\end{center}
\caption{Pseudoplastic features of a two-temperature plasma \eqref{quaccheri} (blue) compared with the Navier-Stokes prediction (dashed). The resummed bulk viscosity coefficient $\zeta_{\text{res}}$ (rescaled by $\zeta_{NS}=2\tau_\Pi (\rho-n)/81$) is plotted as a function of the expansion rate $\nabla_\mu u^\mu$ (in units of $\tau_\Pi^{-1}$).}\label{figunzzuzzuz2}
\end{figure}
\begin{equation}
    \zeta_{\text{res}} = - \dfrac{9{+}\tau_\Pi \nabla_\mu u^\mu {-}3\sqrt{9{+}2\tau_\Pi \nabla_\mu u^\mu {+}(\tau_\Pi \nabla_\mu u^\mu)^2}}{4(\tau_\Pi \nabla_\mu u^\mu)^2} \, 9\zeta_{NS} \, .
\end{equation}
We note that two-temperature plasmas are predominantly pseudoplastic. Appreciable deviations from Navier-Stokes appear when $\tau_\Pi \nabla_\mu u^\mu$ is of order $10$, i.e., 
the nearly-collisionless regime considered in black hole accretion simulations \cite{Chandra:2015iza,Foucart:2017axc}.\\

\noindent\textit{Bulk viscous cosmology.} As the Universe cools down, the relevant degrees of freedom vary, changing the thermodynamic conditions, the degree of ionization, and the radiation-matter ratio \cite{Weinberg_GR_book,Dodelson2003}. Thus, viscous effects in the expanding Universe can be modeled as equilibration processes between the dominant equation of state in a certain era and the dominant equation of state in the next era \cite{Weinberg1971,UdeyIsrael1982}. One can qualitatively describe each transition era-by-era through a simple two-fluid model with four free parameters, which gives us a rough estimate of the pseudoplastic features of the Universe at the transition.

Consider a cosmological fluid comprised of two interacting components, with energy densities $\rho_1$ and $\rho_2$ and pressures $P_1=w_1 \rho_1$ and $P_2=w_2 \rho_2$, with constant $w_1>w_2>0$. The interaction between the two components takes the form of a dissipative energy exchange, which drives the system towards local thermodynamic equilibrium. For clarity, we assume the equilibrium condition is $\rho_1=\alpha \rho_2$, for some constant $\alpha>0$. Then, the conglomerate fluid can be described using the results of this work, with non-equilibrium mode $\phi=\rho_2$, and constitutive relations 
\begin{equation}
    P=w_1 \rho +(w_2{-}w_1)\phi \, , \spc K=(1{+}w_2)\phi \, , \spc F=\bigg[\phi-\dfrac{\rho}{1{+}\alpha} \bigg]C^{-1} \, ,
\end{equation}
where, again, $C>0$ is assumed constant. The exact bulk transport coefficients in the rheological representation are
\begin{equation}
    P_{\text{eq}}= \dfrac{\alpha w_1{+}w_2}{\alpha{+}1}\rho \,  \spc \tau_\Pi =C \, , \spc \dfrac{\zeta}{C}= \dfrac{\rho \alpha(w_1{-}w_2)^2}{(1{+}\alpha)^2}+ \bigg[1+\dfrac{w_1{+}\alpha w_2}{1{+}\alpha} \bigg]  \Pi \, .
\end{equation}
One can solve these equations in an FLRW background \cite{Weinberg_GR_book} with constant Hubble parameter $H{>}0$. The resulting late-time attractor gives us a resummed bulk viscosity coefficient, which can be expressed in terms of  $y=3(w_1{-}w_2)\tau_\Pi H$ as follows:
\begin{figure}
    \centering
    \includegraphics[width=0.5\textwidth]{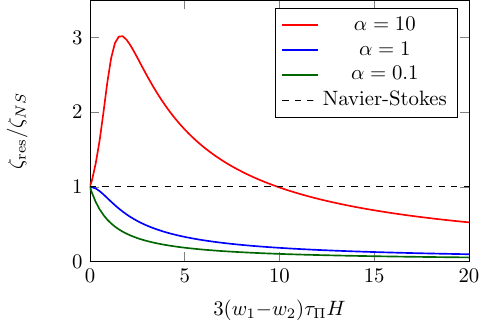}
\caption{Pseudoplastic features of an expanding two-component universe undergoing dissipative energy transfers, as predicted by equation \eqref{pseudouniverse}.}\label{figunzzuzzuz3}
\end{figure}
\begin{equation}\label{pseudouniverse}
    \dfrac{\zeta_{\text{res}}}{\zeta_{NS}} = \dfrac{1{+}\alpha}{2\alpha y^2} \bigg[\alpha (y{-}1) {-}(y{+}1){+}\sqrt{(1{+}\alpha) \big[\alpha (y{-}1)^2+(y{+}1)^2 \big]} \bigg] \, .
\end{equation}
For $\alpha>1$, the Universe exhibits a dilatant behavior for not too large values of $\tau_\Pi H$. For $\alpha \leq 1$, it is always pseudoplastic, namely $\zeta_{\text{res}}<\zeta_{NS}$. Failure to account for these effects may lead to overestimating (or underestimating, in the dilatant case) the impact of viscous effects in the cosmological evolution.\\

\noindent \textit{Conclusions.} We constructed the first causal and stable theory of rheological bulk-viscous systems in relativity. This paves the way for systematically investigating the novel rheological properties displayed by relativistic systems. Our formalism is employed to show that neutron star mergers are intrinsically bulk-viscous systems with rheological pseudoplastic properties. Relativistic pseudoplasticity is also predicted to emerge in two-temperature relativistic plasmas surrounding supermassive galactic black holes. Our framework can be employed to formulate new viscous cosmological models with pseudoplastic features.\\ 


\noindent
\textit{Acknowledgements.}
LG is partially supported by a Vanderbilt Seeding Success Grant. JN is partially supported by the U.S. Department of Energy, Office of Science, Office for Nuclear Physics under Award No. DE-SC0021301. We thank M. Disconzi and E. Most for reading the manuscript and providing useful comments. LG thanks  Andrew Chael for a deeply insightful discussion on two-temperature plasmas, and Gabriel Soares Rocha for interesting conversations about the relationship between rheology and attractors.


\bibliography{Biblio.bib,Biblio2.bib}{}

\newpage

\onecolumngrid
\newpage
\begin{center}
  \textbf{\large Relativistic Bulk Rheology: From Neutron Star Mergers to Viscous Cosmology\\ 
Supplementary Material}\\[.2cm]
  L. Gavassino$^{1}$ and J. Noronha$^2$\\[.1cm]
  {\itshape ${}^1$Department of Mathematics, Vanderbilt University, Nashville, TN, USA\\
  ${}^2$Illinois Center for Advanced Studies of the Universe \& Department of Physics,\\University of Illinois at Urbana-Champaign, Urbana, IL 61801-3003, USA\\}
(Dated: \today)\\[1cm]
\end{center}

\setcounter{equation}{0}
\setcounter{figure}{0}
\setcounter{table}{0}
\setcounter{page}{1}
\renewcommand{\theequation}{S\arabic{equation}}
\renewcommand{\thefigure}{S\arabic{figure}}

\maketitle

\section{Non-linear causality condition for a reactive mixture}

Let us choose as our primary variables the vector $\Psi=(\mathfrak{s},u^\nu,n,Y)^T$. The equations of motion of the mixture are
\begin{equation}\label{sevenuzzo}
    \begin{split}
& u^\alpha \nabla_\alpha \mathfrak{s}= -T^{-1} \mathbb{A}F \, , \\
& (\rho+P)u^\beta \nabla_\beta u_\alpha +G_1\Delta\indices{^\beta _\alpha} \nabla_\beta \mathfrak{s}+G_2\Delta\indices{^\beta _\alpha} \nabla_\beta n+G_3\Delta\indices{^\beta _\alpha} \nabla_\beta Y=0 \, , \\
& u^\alpha \nabla_\alpha n +n \nabla_\alpha u^\alpha =0 \, , \\
& u^\alpha \nabla_\alpha Y =-F \, ,
    \end{split}
\end{equation}
where $\Delta^{\alpha \beta}=g^{\alpha \beta}+u^\alpha u^\beta$ is the orthogonal projector to $u^\mu$, and the coefficients $G_i$ are thermodynamic derivatives, defined though the differential $dP=G_1 \, d\mathfrak{s}+G_2 \, dn +G_3 \, dY$. The system \eqref{sevenuzzo} is in the form $\mathcal{M}^\mu \nabla_\mu \Psi=\mathcal{N}$, with
\begin{equation}
    \mathcal{M}^\mu =
    \begin{bmatrix}
        u^\mu & 0 & 0 & 0 \\
        G_1 \Delta^{\alpha \mu} & (\rho+P) \delta^\alpha_\nu u^\mu & G_2 \Delta^{\alpha \mu} & G_3 \Delta^{\alpha \mu} \\
        0 & n\delta^\mu_\nu & u^\mu & 0 \\
        0 & 0 & 0 & u^\mu \\
    \end{bmatrix} .
\end{equation}
Therefore, we can follow the same procedure as in \cite{Causality_bulk}, and compute the characteristic determinant $\mathcal{P}(\xi)=\det(\mathcal{M}^\mu \xi_\mu)$, for a generic covector $\xi_\mu$. The result is
\begin{equation}\label{expectopatronum}
    \mathcal{P}(\xi)= (u^\alpha \xi_\alpha)^5 (\rho+P)^4 \bigg[ (u^\mu \xi_\mu)^2- \dfrac{nG_2}{\rho+P} \Delta^{\mu \nu}\xi_\mu \xi_\nu \bigg] \, .
\end{equation}
Invoking the same reasoning as in \cite{Causality_bulk}, we can conclude that the characteristic speed of sound squared is
\begin{equation}
    c_{\text{char}}^2 = \dfrac{nG_2}{\rho+P}  \, .
\end{equation}
This quantity can be rewritten in a more familiar form if we consider that
\begin{equation}
  \dfrac{n}{\rho+P}= \dfrac{\partial n}{\partial \rho}\bigg|_{\mathfrak{s},Y} \, , \spc  G_2 = \dfrac{\partial P}{\partial n}\bigg|_{\mathfrak{s},Y} \, , 
\end{equation}
giving the expected result:
\begin{equation}
    c_{\text{char}}^2  = \dfrac{\partial P}{\partial \rho}\bigg|_{\mathfrak{s},Y}  \, .
\end{equation}
For the fluid mixture to be causal, one needs to have $c_\text{char}^2\in [0,1]$, which is true of our test fluid, see equation \eqref{oureossound}.

\newpage
\section{Consistency of our toy model for neutron-star matter}

Given the function $\rho(n,\mathfrak{s},Y)$, where $\rho>0$ is the energy density, $n>0$ is the conserved particle density, $\mathfrak{s}>0$ is the entropy per conserved particle, and $Y$ is the non-conserved fraction, the first law of thermodynamics gives 
\begin{equation}\label{drho}
    d\rho = \dfrac{\rho+P}{n} \, dn +Tn \, d\mathfrak{s}-\mathbb{A}n \, dY \, ,
\end{equation}
where $P$ is the total (non-equilibrium) pressure, $T$ is the non-equilibrium temperature\footnote{Note that the temperature defined in \eqref{drho} does not coincide with the Israel-Stewart temperature \cite{Israel_Stewart_1979}, which is just $T(\Pi=0)$.}, and $\mathbb{A}$ is the affinity of any reaction involving the creation of a particle of type ``$Y$''. Thus, if the equation of state is 
\begin{equation}\label{rhorho}
    \rho = (1+\mathfrak{s}^2+e^{Y})n + n^2 e^{-Y} \, ,
\end{equation}
we have that $P=n^2 e^{-Y}>0$, $T=2\mathfrak{s}>0$, and $\mathbb{A}=ne^{-Y}-e^Y$. The condition for chemical equilibrium is $\mathbb{A}=0$, which implies that $Y_{\text{eq}}(\rho,n)=\ln\sqrt{n}$. It is evident that the minimum energy principle is respected (namely, $Y_{\text{eq}}$ minimizes $\rho$ for fixed $\mathfrak{s}$ and $n$ \cite{Callen_book}), since $\partial^2_Y \rho(n,\mathfrak{s},Y)=n(e^Y+ne^{-Y})>0$.
The characteristic speed of sound is\footnote{The fact that the signal propagation is determined by the speed of sound \textit{at constant fractions} is well known \cite{Camelio2022}. Nevertheless, we provided a direct proof in the previous section.}
\begin{equation}\label{oureossound}
    c_{\text{char}}^2 = \dfrac{\partial P}{\partial \rho}\bigg|_{\mathfrak{s},Y} = \dfrac{2ne^{-Y}}{1+\mathfrak{s}^2+e^Y+2ne^{-Y}} \, .
\end{equation}
Recalling that $n>0$, we have $c_{\text{char}}^2 \in (0,1)$, meaning that the fluid is causal arbitrarily far from equilibrium (i.e. for all values of $Y$). Also, since $\rho+P>0$, it follows that the fluid equations are strongly hyperbolic \cite{GerochLindblom1990,Causality_bulk}.

To prove that the equation of state \eqref{rhorho} is thermodynamically consistent, it is enough to note that the Hessian of the specific energy is positive definite (see \citet{landau5}, \S 96, ``\textit{Thermodynamic inequalities for solutions}''):
\begin{equation}
    H_{v\rho} = 
\begin{bmatrix}
-\dfrac{\partial P}{\partial v}\bigg|_{\mathfrak{s},Y} & -\dfrac{\partial P}{\partial \mathfrak{s}}\bigg|_{v,Y} & -\dfrac{\partial P}{\partial Y}\bigg|_{v,\mathfrak{s}} \\ 
\textcolor{white}{+}\dfrac{\partial T}{\partial v}\bigg|_{\mathfrak{s},Y} & \textcolor{white}{+}\dfrac{\partial T}{\partial \mathfrak{s}}\bigg|_{v,Y} & \textcolor{white}{+}\dfrac{\partial T}{\partial Y}\bigg|_{v,\mathfrak{s}} \\ 
-\dfrac{\partial \mathbb{A}}{\partial v}\bigg|_{\mathfrak{s},Y} & -\dfrac{\partial \mathbb{A}}{\partial \mathfrak{s}}\bigg|_{v,Y} & -\dfrac{\partial \mathbb{A}}{\partial Y}\bigg|_{v,\mathfrak{s}} \\ 
\end{bmatrix} =
\begin{bmatrix}
2n^3 e^{-Y} & 0 & n^2 e^{-Y} \\ 
0 & \, 2 \, & 0 \\ 
n^2 e^{-Y} & 0 & e^Y +ne^{-Y} \\ 
\end{bmatrix} ,
\end{equation}
where $v=n^{-1}$ is the specific volume \cite{ChoquetBruhatGRBook}. In particular, we have that $\det (H_{v\rho}) = 4n^3+2n^4 e^{-2Y}>0$.

Let us now focus on the returning force $F$, which is postulated to be
\begin{equation}
   F=(n^{-1/2}e^Y-1)C^{-1}, 
\end{equation}
where $C>0$ is a constant. Recalling that the condition for chemical equilibrium is $e^{Y_{\text{eq}}}{=}\sqrt{n}$, we have that $F(n,Y_{\text{eq}})=0$, as it should be. Furthermore, if $e^Y{>}\sqrt{n}$, then $u^\mu \nabla_\mu Y{=}{-}F{<} 0$, while if $e^Y{<}\sqrt{n}$, then $u^\mu \nabla_\mu Y{=}{-}F {>} 0$, meaning that the chemical reaction drives $Y$ towards equilibrium. Finally, if we apply ``$u^\mu \nabla_\mu$'' on both sides of \eqref{rhorho}, and we invoke the field equations $u^\mu \nabla_\mu \rho=-(\rho+P)\nabla_\mu u^\mu$, $u^\mu \nabla_\mu n=-n\nabla_\mu u^\mu$, and $u^\mu \nabla_\mu Y=-F$, we obtain the following:
\begin{equation}\label{entropiesco}
    T\nabla_\mu (n\mathfrak{s}u^\mu)= \dfrac{P}{C} \bigg(1+\dfrac{e^Y}{\sqrt{n}}\bigg) \bigg( 1- \dfrac{e^{Y}}{\sqrt{n}}\bigg)^2 \geq 0 \, .
\end{equation}
As can be seen, the second law of thermodynamics holds arbitrarily far from equilibrium. Combining this fact with (a) the positive definiteness of $H_{v\rho}$, (b) the inequality $\rho+P>0$, and (c) the observation that the dynamics is causal, we can then invoke the Gibbs stability criterion \cite{GavassinoGibbs2021,GavassinoStabilityCarter2022}, together with Theorem II of \cite{GavassinoSuperluminal2021}, and conclude that the fluid is stable, both thermodynamically and hydrodynamically \cite{GavassinoCausality2021}. This completes our consistency check.

\newpage
\section{Israel-Stewart as perturbative rheology}

\subsection{DNMR and Hiscock-Lindblom and rheological models}
The alternative versions of the IS theory currently in use in the literature arise from different prescriptions for $\tau_\Pi(\rho,n,\Pi)$ and $\zeta(\rho,n,\Pi)$ in our rheological model, $\tau_\Pi(\Pi) \, \dot{\Pi}+\Pi=-\zeta(\Pi) \, \theta$. For example, setting
\begin{equation}
\tau_\Pi = \dfrac{\tau_0(\rho,n)}{1+\lambda(\rho,n)\Pi} , \quad
\zeta= \dfrac{\zeta_0(\rho,n)+\nu(\rho,n)\Pi}{1+\lambda(\rho,n)\Pi},
\end{equation}
Eq. (7) of the main text becomes DNMR theory \cite{DMNR2012}
\begin{equation}
\tau_0 u^\mu \nabla_\mu \Pi+\Pi+\lambda \Pi^2 =-\zeta_0 \nabla_\mu u^\mu -\nu \Pi \nabla_\mu u^\mu \, .
\end{equation}
If, instead, we set
\begin{equation}\label{Hisck}
\tau_\Pi= \zeta_0\beta_0 ,\qquad 
\zeta= \zeta_0{+}\dfrac{\zeta_0 T \Pi}{2}\bigg[\dfrac{\beta_0}{T} {-}(\rho{+}P_{\text{eq}}{+}\Pi)\dfrac{\partial (\beta_0/T)}{\partial\rho}\bigg|_n \! \! {-}n\dfrac{\partial(\beta_0/T)}{\partial n}\bigg|_\rho \bigg], 
\end{equation}
where $T(\rho,n)$ is the equilibrium temperature, and $\zeta_0(\rho,n)$ and $\beta_0(\rho,n)$ are two positive functions, one recovers Hiscock-Lindblom theory \cite{Hishcock1983}
\begin{equation}\label{HLLL}
\Pi= -\zeta_0 \bigg[ \nabla_\mu u^\mu +\beta_0 u^\mu \nabla_\mu \Pi +\dfrac{\Pi T}{2} \nabla_\mu \bigg( \beta_0 \dfrac{u^\mu}{T} \bigg) \bigg].
\end{equation}

\subsection*{Is Hiscock-Lindblom really better than Maxwell-Cattaneo?} 

In its original (near-equilibrium) formulation, IS theory arises from a second-order expansion of the entropy density and production rate for small values of $\Pi$ \cite{Israel_Stewart_1979}, leading to the Hiscock-Lindblom equation \eqref{HLLL}. Usually, in applications, one neglects the term proportional to $\nabla_\mu (\beta_0 u^\mu/T)$, assuming it to be of higher order than the term $\nabla_\mu u^\mu$. This results in the Maxwell-Cattaneo (MC) theory \cite{Zakari1993}. However, some authors have argued \cite{Salmonson1991,Zakari1993,Maartens1995,
Zimdahl1996} that the term $\nabla_\mu (\beta_0 u^\mu/T)$ is required for thermodynamic consistency, and it should not be neglected, despite being formally different from the corresponding term $\nu \Pi \nabla_\mu u^\mu$ in DNMR theory \cite{DMNR2012}. There is no strong consensus on whether this term must be included or not in numerical simulations \cite{Dore:2020jye,Chab2021,Camelio2022}. Our results can settle this debate.

Since our model for neutron-star bulk viscosity is thermodynamically consistent arbitrarily far from chemical equilibrium, Hiscock-Lindblom theory should arise as an approximation of this model for small $\Pi$. To second order in $\Pi$, equation \eqref{rhorho} in this note becomes
$\rho = (1+\mathfrak{s}^2)n+2n^{3/2}+n^{-3/2}\Pi^2 +\mathcal{O}(\Pi^3)$,
and the Hiscock-Lindblom transport coefficients are \footnote{In the computation of $\beta_0$, we have invoked the duality between the maximum entropy principle and the minimum energy principle, see \citet{Callen_book}, section 5-1.}
\begin{equation}
    \zeta_0= \zeta(\Pi=0)= \dfrac{n^{3/2}}{2}C , \quad
\beta_0= -T \dfrac{\partial^2 (n\mathfrak{s})}{\partial \Pi^2}\bigg|_{\rho,n}=\dfrac{\partial^2 \rho}{\partial \Pi^2}\bigg|_{\mathfrak{s},n} = 2n^{-3/2}, \qquad
T= \bigg[\dfrac{\partial \rho(\Pi=0)}{\partial (n \mathfrak{s})}\bigg|_{n}\bigg]^{-1}=2\sqrt{\dfrac{\rho-n-2n^{3/2}}{n}}. 
\end{equation}
Plugging them into \eqref{Hisck}, we obtain
    $\tau_\Pi = C$, 
    $\zeta= \bigg[\dfrac{n^{3/2}}{2} +\dfrac{5}{4}\Pi + \dfrac{\Pi^2}{4(\rho-n-2n^{3/2})} \bigg]C$.
Comparing the latter formula for $\zeta(\rho,n,\Pi)$ with the ``exact'' formula (10) in the main text, we see that the linear correction in $\Pi$ coming from the term $\propto \nabla_\mu (\beta_0 u^\mu/T)$ has an incorrect prefactor $5/4$, instead of the ``exact'' factor $2$. This shows that the term $\propto \nabla_\mu (\beta_0 u^\mu/T)$ is of the same order as the truncation error that we commit when we expand the theory for small $\Pi$. Thus, this term is spurious, and it should not be trusted (in the next subsection, we provide an intuitive explanation for this fact). Given that such spurious term further complicates numerical implementations \cite{CamelioSimulations2022}, it is reasonable to favor MC (which is simpler at the given order) or DNMR (which is more refined) over the Hiscock-Lindblom formulation.

\subsection*{The problem with the Hiscock-Lindblom truncation}

Let us briefly review the original derivation of the Israel-Stewart theory by \citet{Hishcock1983}. First, one postulates the existence of an entropy current, which is truncated to second order in $\Pi$:
\begin{equation}\label{smuuu}
    s^\mu = \bigg( n\mathfrak{s} - \dfrac{\beta_0 \Pi^2}{2T} \bigg)u^\mu \, .
\end{equation}
Here, $\mathfrak{s}$, $T$, and $\beta_0$ are functions of $\rho$ and $n$ (and not of $\Pi$). Then, the divergence of $s^\mu$ is computed and, with the aid of the conservation laws $\nabla_\mu T^{\mu \nu}=0$ and $\nabla_\mu n^\mu=0$, it is recast in the following form (see \cite{Hishcock1983} for the details):
\begin{equation}\label{gameto1}
    T\nabla_\mu s^\mu = -\Pi \bigg[\nabla_\mu u^\mu +\beta_0 u^\mu \nabla_\mu \Pi +\dfrac{\Pi T}{2} \nabla_\mu\bigg(\dfrac{\beta_0 u^\mu }{T}\bigg) \bigg].
\end{equation}
Finally, one imposes the validity of the second law by requiring that $T\nabla_\mu s^\mu =\Pi^2/\zeta_0$, where $\zeta_0$ is also a function of $\rho$ and $n$ only. This produces the field equation
\begin{equation}\label{azkaban}
    \Pi = -\zeta_0 \bigg[\nabla_\mu u^\mu +\beta_0 u^\mu \nabla_\mu \Pi +\dfrac{\Pi T}{2} \nabla_\mu\bigg(\dfrac{\beta_0 u^\mu }{T}\bigg) \bigg] .
\end{equation}
The term proportional to $\nabla_\mu (\beta_0 u^\mu/T)$ appears to be a necessary ingredient for the second law of thermodynamics to hold, but it is a mere byproduct of the second-order truncation in $\Pi$, and it disappears if we increase the truncation order. To understand this point, consider a hypothetical higher-order extension of the theory, such that
\begin{equation}\label{gameto2}
    T \nabla_\mu s^\mu = \dfrac{\Pi^2}{ \zeta_0 f},
\end{equation}
where $f(\rho,n,\Pi)=1+a(\rho,n)\Pi+...$ is a positive function. With this generalized choice of entropy production rate, assuming for simplicity that the entropy current is still given by equation \eqref{smuuu} (so that \eqref{gameto1} still holds), we obtain the following higher-order field equation:
\begin{equation}
    \Pi = -\zeta_0 \bigg[\nabla_\mu u^\mu +\beta_0 u^\mu \nabla_\mu \Pi +\dfrac{\Pi T}{2} \nabla_\mu\bigg(\dfrac{\beta_0 u^\mu }{T}\bigg)+a\Pi \nabla_\mu u^\mu+ a\beta_0 \, \Pi \, u^\mu \nabla_\mu \Pi +...\bigg] .
\end{equation}
We immediately see that the term $a\Pi \nabla_\mu u^\mu$ is of the same order as the term proportional to $\nabla_\mu (\beta_0 u^\mu/T)$. Indeed, we can even choose $a$ such that these two terms cancel out. The implication is simple: the last term in \eqref{azkaban} has the same size as the truncation error that one makes while approximating the entropy production rate as $\Pi^2/\zeta_0$.

\newpage
\section{Two-temperature plasmas}

\subsection*{Outline of the model}
We consider a neutral high-temperature plasma comprised of an ideal gas of protons and an ideal gas of electrons, which move as a single fluid (i.e. have the same flow velocity $u^\mu$). The conglomerate energy density and pressure are assumed to be the sum of the proton and electron contributions: $\rho=\rho_p+\rho_e$, $P=P_p+P_e$. Since both gases have high temperatures, they are non-degenerate, and we can neglect the Coulomb contribution to the equation of state. Furthermore, we assume that the temperature is so high that the electrons are ultrarelativistic, while the protons are still assumed non-relativistic. Thus, we have the following relations:
\begin{equation}
    \begin{split}
\rho_p ={}& n+\dfrac{3}{2} nT_p \, , \\
P_p ={}& nT_p \, , \\
\rho_e ={}& 3nT_e \, , \\
P_e ={}& nT_e \, , \\
    \end{split}
\end{equation}
where we have set the mass of the proton to $1$. Note that the proton and electron particle densities are both equal to $n$, because the plasma is assumed neutral. On the other hand, the kinetic temperatures are in general different, because the mean inelastic $pe$ collision rate is much smaller than the mean inelastic $pp$ and $ee$ collision rates, so that the $p$ and the $e$ gas achieve kinetic equilibrium \textit{separately} before reaching thermal equilibrium with each other. The heat exchange between the protons and electrons due to inelastic $pe$ collisions is $Q=u_\nu \nabla_\mu T^{\mu \nu}_p=-u_\nu \nabla_\mu T^{\mu \nu}_e$, and we can express it in the form $Q=-C^{-1}n(T_e-T_p)$, where $C>0$ is a transport coefficient\footnote{In general, $C$ is a function of $n$, $T_p$ and $T_e$. Here, we will take it to be a constant, for simplicity.}. This results in a simple equation of motion for the electron pressure:
\begin{equation}\label{pippuzzuz}
    3 u^\mu \nabla_\mu P_e =-4P_e \nabla_\mu u^\mu -(P_e{-}P_p)C^{-1} \, .
\end{equation}

\subsection*{General formulation}

Since the thermodynamic state of the plasma is uniquely characterized by $\{n,T_p,T_e\}$, we can change variables to $\{\rho,n,\phi\}$, for some choice of $\phi$, and recast the model in the standard form used in the main paper. Taking $\phi=P_e$, we obtain
\begin{equation}
    \begin{split}
\rho_p ={}& \rho-3\phi \, , \\
P_p ={}& \dfrac{2}{3} (\rho-n)-2\phi \, , \\
\rho_e ={}& 3\phi \, , \\
P_e ={}& \phi \, . \\
    \end{split}
\end{equation}
From this, we immediately obtain $P =\frac{2}{3}(\rho-n)-\phi$. Furthermore, equation \eqref{pippuzzuz} becomes
\begin{equation}
    u^\mu \nabla_\mu \phi = -\dfrac{4}{3} \phi \nabla_\mu u^\mu -\bigg[\phi -\dfrac{2}{9}(\rho-n) \bigg]C^{-1} \, ,
\end{equation}
which is what we wanted to prove.

\newpage

\label{lastpage}

\end{document}